\documentclass[%
 reprint,
superscriptaddress,
 amsmath,amssymb,
 aps,
]{revtex4-2}

\usepackage{graphicx}
\usepackage{dcolumn}
\usepackage{bm}
\usepackage[mathlines]{lineno}
\usepackage{xcolor}

\begin{document}

\preprint{APS/123-QED}

\title{The Interplay Between Solute Atoms and Vacancy Clusters in Magnesium Alloys}

\author{Peng Yi}
\affiliation{Department of Materials Science and Engineering, Johns Hopkins University, Baltimore, MD 21218, USA}
\affiliation{Hopkins Extreme Materials Institute, Johns Hopkins University, Baltimore, MD 21218, USA}

\author{Taisuke Sasaki}
\affiliation{National Institute for Materials Science, 1-2-1 Sengen, Tsukuba, Ibaraki 305-0047, JAPAN}
\affiliation{Elements Strategy Initiative for Structural Materials (ESISM), Kyoto University, Kyoto 606-8501, JAPAN}

\author{Suhas Eswarappa Prameela}
\affiliation{Department of Materials Science and Engineering, Johns Hopkins University, Baltimore, MD 21218, USA}
\affiliation{Hopkins Extreme Materials Institute, Johns Hopkins University, Baltimore, MD 21218, USA}


\author{Timothy P. Weihs}
\affiliation{Department of Materials Science and Engineering, Johns Hopkins University, Baltimore, MD 21218, USA}
\affiliation{Hopkins Extreme Materials Institute, Johns Hopkins University, Baltimore, MD 21218, USA}
\affiliation{Department of Mechanical Engineering, Johns Hopkins University, Baltimore, MD 21218, USA}

\author{Michael L. Falk}
\affiliation{Department of Materials Science and Engineering, Johns Hopkins University, Baltimore, MD 21218, USA}
\affiliation{Hopkins Extreme Materials Institute, Johns Hopkins University, Baltimore, MD 21218, USA}
\affiliation{Department of Mechanical Engineering, Johns Hopkins University, Baltimore, MD 21218, USA}
\affiliation{Department of Physics and Astronomy, Johns Hopkins University, Baltimore, MD 21218, USA}


\date{\today}

\begin{abstract}
Atomic-scale calculations indicate that both stress effects and chemical binding contribute to the redistribution of solute in the presence of vacancy clusters in magnesium alloys.  As the size of the vacancy cluster increases, chemical binding becomes more important relative to stress.  By affecting the diffusivity of vacancies and vacancy clusters, solute atoms facilitate clustering and stabilize the resulting vacancy clusters, increasing their potential to promote solute segregation and to serve as heterogeneous nucleation sites during precipitation. Experimental observation of solute segregation in simultaneously deformed and aged Mg-Al alloys provides support for this mechanism.
\begin{description}
\item[Keywords]
processing, precipitation, deformation, defects
\end{description}
\end{abstract}

\maketitle



Vacancies are crystal lattice defects that occur spontaneously in equilibrium at finite temperatures, the populations of which can be substantially enhanced through fast quenching, irradiation, or severe plastic deformation (SPD) \cite{PorterEsterling2009}.  They play particularly important roles in the intermetallic precipitation process in metallic alloys.  Their role in accelerating the kinetics of precipitation through vacancy-enhanced diffusion is well recognized~\cite{Wu2016,DESCHAMPS2021} and is exploited to control the precipitation process~\cite{Pogatscher2014,Sun2019,Zhu2021}.  

On the other hand, the thermodynamic roles vacancies play in the nucleation of precipitates have not drawn significant attention. As early as the 1970s~\cite{Katz1981}, it was proposed that vacancies may also form clusters that serve as heterogeneous nucleation sites for precipitation, like the well-established role played by dislocations. However, this has not been demonstrated convincingly since direct imaging of vacancy clusters remains challenging.  One of the counter-arguments is the relative short lifetime of single vacancies.  Militzer et al.~\cite{Militzer1994} presented a model to account for the thermodynamic effects of deformation-induced single vacancies on precipitate nucleation. According to their calculation, the lifetimes of single vacancies are not likely to be long enough for them to play a major role thermodynamically. This, however, neglects the fact that single vacancies can aggregate into vacancies clusters that have longer lifetimes. Experimental methods like positron annihilation spectroscopy (PAS) have provided indirect evidence of the existence of clusters of several dozen vacancies~\cite{CampilloRobles2007,Liu2016,Cizek2018}.  Vacancy clustering in Al was also demonstrated in some computational studies~\cite{WangPRB2011,Adibi2020}.  Some recent studies also provide evidence of solute segregation near vacancies and vacancy clusters in Al alloys~\cite{Berg2001,Sun2019}.  These solute segregation can significantly reduce the barrier to intermetallic nucleation.

We are interested in investigating the interactions between solute atoms and vacancy clusters in Mg alloys because Mg is a lightweight material with great potential and increasing interest.  Recent experiments have demonstrated that the precipitate density and morphology in Mg alloys can be significantly improved by deformation processing \cite{Ma2019,Prameela2020,PRAMEELA20201476,Prameela2020c}, which could be attributed to the excess vacancies generated during SPD, as demonstrated computationally~\cite{Yi2017a,Yi2021}.  During traditional processing, vacancies in Mg can diffuse to sinks rapidly because basal dislocations are typically distributed uniformly, as compared to the cellular dislocation sub-structures in Al~\cite{Cizek2018,Cizek2019}.  However, experimental evidence exists that SPD could also create cellular sub-structure in Mg alloys which could allow vacancies to segregate into clusters, as seen in Al alloys~\cite{Li2008,Lee2018}.  

In this study, we use computer simulation to demonstrate solute segregation due to small vacancy clusters in Mg alloys, analyze its underlying mechanism, and examine the stability of vacancy clusters in solid solutions.  To validate this hypothesis, we perform equal channel angular extrusion (ECAE) processing on the Mg-Al binary alloy to generate a high density of vacancies and vacancy clusters.  ECAE is performed at a typical aging temperature to allow co-evolution of solutes and vacancies.  The solute segregation in the simultaneously deformed and aged samples characterized using 3D atom probe tomography (APT) are used to provide essential support for the mechanism of vacancy clusters induced solute segregation.

\begin{figure*}
\includegraphics[width=1.0\textwidth]{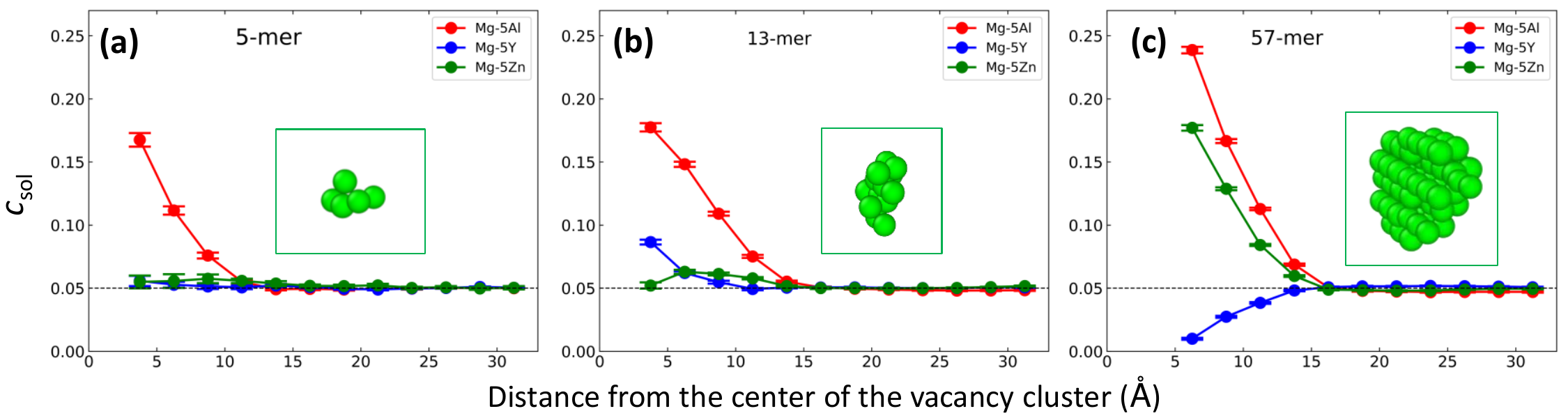}
\caption{\label{fig:segregation} Solute concentration as a function of the distance from the center of a vacancy cluster at 450K.  Vacancy clusters are a (a) 5-mer, (b) 13-mer, or (c) 57-mer.  Mg-5X represents a binary Mg-X solid solution with 5 at.\% solute.  Error bars represent standard errors, and lines are guides to the eye.  Insets: contours of the radial profile of solute concentration for (a) Mg-5Al near a 5-mer, (a) Mg-5Zn near a 13-mer, and (c) Mg-5Y near a 57-mer.  The results are calculated using MEAM potentials with zero external pressure.  Insets are snapshots of the vacancy clusters.}
\end{figure*}



Molecular dynamics (MD) and Monte Carlo (MC) simulations were performed using the LAMMPS package~\cite{Plimpton1995}.  Three Mg-X binary alloy systems (X=Al, Y, Zn) were modeled using the MEAM potentials~\cite{Kim2009,Kim2015,Jang2018}, which were used  in our previous studies~\cite{Yi2016, Yi2017a, Yi2019, Yi2021}. We chose these three binary systems because of the widespread use of these elements in alloying Mg for commercial use.  The contrast between these three solutes also offer great benefits for comparative studies.  We additionally deployed density functional theory (DFT) calculations to validate the accuracy of the predictions made with these MEAM potentials.  The DFT calculations were performed using the VASP package~\cite{Kresse1996a, Kresse1996b}  (See supplementary materials for computation and experiment details).


Experiments were performed on Mg-Al alloys. To start with a precipitate-free and well-controlled material, an as-cast binary Mg-9wt.\%Al ingot was subjected to solution-treatment at 450\textdegree C for 24 hours with a protective argon gas flow, followed by quenching in ice water to avoid precipitation.  Rectangular-shaped billets with a size of 6.35$\times$6.35$\times$19 mm$^3$ were cut from the center of the solution-treated sheet for ECAE processing.  The cut samples were subjected to the one-pass ECAE at 150\textdegree C with an extrusion rate of 0.15 mm/min and a backpressure of 0.45 MPa.  Needle-shaped samples were then obtained from ECAE'ed samples for APT mapping.

%
%


Using the hybrid MC/MD simulations, the equilibrium solute distributions near vacancy clusters are shown in Fig.~\ref{fig:segregation} for Mg-Al, Mg-Y, and Mg-Zn systems at 450K, which is the typical processing temperature for Mg alloys.  The vacancy clusters consist of either 5, 13, or 57 vacancies, termed 5-mer, 13-mer, and 57-mer, respectively.  If spherical, they have radii of about 3, 4, and 7 $\mathrm{\AA}$, respectively.

Near 5-mer and 13-mer (Fig.~\ref{fig:segregation}(a)(b)), solute segregation is observed for all three alloys, and it is the most significant in Mg-Al.  The Mg-Zn system shows a non-monotonic solute concentration change, which will be discussed later.  Near a larger vacancy cluster, 57-mer (Fig.~\ref{fig:segregation}(c)), the segregation becomes more significant for Mg-Al and  Mg-Zn.  However, solute depletion occurs in Mg-Y.

Vacancy and vacancy clusters generate stress fields that lead to solute redistribution due to the reduction of the strain energy.  On the other hand, different chemical binding tendency between solute and vacancies could also significantly contribute to solute redistribution.  In the following, we assess the relative importance of these two mechanisms.

\begin{figure*}
\includegraphics[width=1.0\textwidth]{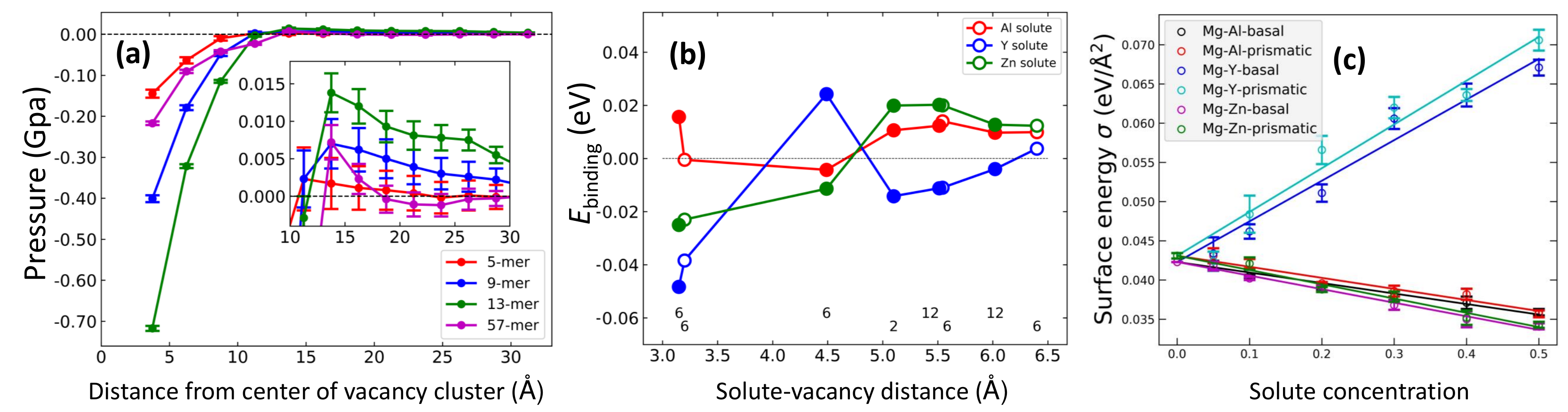}
\caption{\label{fig:causes} Results calculated using MEAM potentials. (a) Local hydrostatic pressure as a function of the distance from the center of a vacancy cluster in pure Mg at 450K with zero external pressure.  Vacancy clusters are 5-mer, 9-mer, 13-mer, or 57-mer.  Error bars represent standard errors, and lines are guide to the eye.  Inset: enlarged area showing the ``shoulder" of the pressure profile.  (b) Solute-vacancy binding energy as a function of the solute-vacancy pair distance.  Different data points represent solute atom at different nearest neighbor (NN) and second nearest neighbor (2NN) sites.  There are 2 unique NN sites and 6 unique 2NN sites.  Numbers below the symbols represent the number of their equivalent sites.  The distance is calculated using a hexagonal lattice with lattice constants $a=3.2\mathrm{\AA}$ and $c=5.1\mathrm{\AA}$.  Open circles represent that solute and vacancy are on the same basal plane, and solid circles represent that solute and vacancy are on different basal planes. Lines are guide to the eye. (c) Surface energies at 0K as a function of solute concentration for Mg-Al, Mg-Y, and Mg-Zn systems.  Lines are linear fits.}
\end{figure*}

To evaluate the significance of the strain energy effect, we first calculated the hydrostatic pressure near a vacancy cluster as shown in Fig.~\ref{fig:causes}(a).  A hydrostatic tension field arises even when no external hydrostatic pressure is applied.  With increasing vacancy cluster size, the tension first increases then decreases.  This tension field, as well as a ``shoulder" in the profile, has also been experimentally observed in colloidal systems~\cite{Lin2016}.  It is worth noting that linear isotropic elastic theory predicts that the hydrostatic pressure near a void is identical to the applied hydrostatic pressure~\cite{He2006}, and nonlinear elasticity is needed to capture this hydrostatic pressure~\cite{Lin2016}.

\begin{table}[h]
\caption{\label{tab:atomicvolume}%
Relative atomic size change, \textrm{$\Delta v/v$}, of three different solute atoms in Mg matrix.}
\begin{ruledtabular}
\begin{tabular}{lddd}
\multicolumn{1}{c}{Solute}&
\multicolumn{1}{c}{\textrm{MEAM}}&
\multicolumn{1}{c}{\textrm{DFT}}&
\multicolumn{1}{c}{\textrm{EXP}}\\
\colrule
Al & -0.30 & -0.39 & -0.358\\
Y & 0.24 & 0.44 & \mathrm{N/A}\\
Zn & -0.32 & -0.53 & -0.488\\
\end{tabular}
\end{ruledtabular}
\end{table}

A hydrostatic tension field should favor solute atoms larger than the Mg solvent atoms.  We calculated the solute atom size in the Mg matrix, $\Delta v/v$, where $\Delta v$ is the volume size change due to one substitutional solute atom, and $v$ is the average atomic volume in a pure Mg crystal.  The MEAM and DFT calculation results are summarized in Table~\ref{tab:atomicvolume}, as well as the experimental measurements~\cite{King1966}.  The MEAM and DFT predictions are consistent and in good agreement with available experimental data.  The atom sizes in descending order is Y$>$Mg$>$Al$>$Zn, in the Mg matrix.  Therefore, reductions in elastic strain energy will drive Al and Zn depletion and Y segregation near vacancy clusters.  However, only the Y segregation near 5-mer and 13-mer in Fig.~\ref{fig:segregation} follows this prediction, suggesting that the other mechanism, chemical binding, are also important in determining the solute distribution.


We calculated the solute-vacancy binding energy at 0K using the MEAM potentials, as shown in Fig.~\ref{fig:causes}(b), where positive values represent attraction.  Based on Fig.~\ref{fig:causes}(b), a vacancy has an  attractive binding with an Al solute atom at most of the NN and 2NN sites, which is the complete opposite for a Y solute atom.  A vacancy has a negative binding with a Zn solute atom for NN sites; however, the attraction between a vacancy and a Zn atom at 2NN sites can still lead to Zn segregation near vacancy clusters.  The change of sign of solute-vacancy binding energy at different neighboring sites for Zn solute is likely the reason for the non-monotonic solute concentration profile for Mg-5Zn observed in Fig.~\ref{fig:segregation}(a) and (b).

We compared the MEAM predictions of the binding energies for the NN sites to the experiments, as shown in Table~\ref{tab:ebinding}.  The experimental binding energies are much greater in magnitude than MEAM results.  Assuming that the NN sites dominate the solute distribution near small vacancy clusters, this difference suggests that the solute segregation in real experiments to be more significant than that observed using these MEAM potentials. 

We also calculated the binding energies using DFT for comparison (Table~\ref{tab:ebinding}).  Our DFT calculations are in good agreement with previous work~\cite{Shin2010,Zhou2016}.  The MEAM and DFT are most consistent for the Y solute.  Both methods predict a largely negative binding energy between a vacancy and a Y solute at NN sites.  The MEAM predicts weaker solute-vacancy binding for Al solute at NN sites, as compared to DFT.  The MEAM also predicts a repulsion between Zn and a vacancy at NN sites, while the DFT predicts an attraction.   However, it is known that the DFT calculation underestimates the Al-vacancy binding energy and predicts that Al has a weaker binding with a vacancy than Zn, although experimentally the opposite is measured (Table~\ref{tab:ebinding})~\cite{Shin2010,Zhou2016}.

\begin{table}[h]
\caption{\label{tab:ebinding}%
Solute-vacancy binding energy, $E_{\mathrm{binding}}$ (eV), for nearest-neighbor (NN) sites.  The values are averaged over all NN sites on and off the basal plane.}
\begin{ruledtabular}
\begin{tabular}{llll}
\multicolumn{1}{c}{\textrm{Solute}}&
\multicolumn{1}{c}{\textrm{MEAM}}&
\multicolumn{1}{c}{\textrm{DFT}}&
\multicolumn{1}{c}{\textrm{EXP}}\\
\colrule
Al & 0.0076 & 0.023 (this study) & 0.29$\pm$0.02\cite{Rao1978}\\
  & & 0.03\cite{Shin2010}, 0.05\cite{Zhou2016} & \\
Y & -0.043 & -0.060 (this study) & N/A\\
  & & -0.07\cite{Shin2010}, -0.065\cite{Zhou2016} & \\
Zn & -0.024 & 0.044 (this study) & 0.07$\pm$0.02\cite{Vydyanath1968}\\
  & & 0.05\cite{Shin2010}, 0.045\cite{Zhou2016} & \\
\end{tabular}
\end{ruledtabular}
\end{table}


While the solute-vacancy binding energy is useful in explaining the interaction between a single vacancy and a single solute, the surface energy is a more suitable measure for larger vacancy clusters, for example, a 57-mer.  To test if a larger void might display different solute segregation behavior, we calculate the surface energy at 0K for Mg solid solutions using the MEAM potentials.  

The surface energies of the two lowest index planes as functions of solute concentration are shown in Fig.~\ref{fig:causes}(c).  As predicted by the MEAM potential, the surface energies for pure Mg are 0.042 and 0.043 $\mathrm{eV/\AA^2}$ for the basal plane and prismatic plane, respectively.  DFT predictions are 0.034 and 0.038 $\mathrm{eV/\AA^2}$, for the same planes~\cite{Tran2016}; and the experimental estimate is 0.049 $\mathrm{eV/\AA^2}$ for the basal plane~\cite{TYSON1977,Keene1993,Vitos1998,Kim2009}. The MEAM predictions are closer to the experiments than DFT.

According to Fig.~\ref{fig:causes}(c), Al solute and Zn solute reduce the surface energies of the solid solution, and Y solute increases the surface energies.  Therefore, the surface energy change due to the presence of solute atoms favors the incorporation of Al and Zn solute into void surfaces and rejection of Y solute, which is consistent with the solute distribution near 57-mer(Fig.~\ref{fig:segregation}(c)).  It also shows that the NN sites binding no longer dominates the Zn solute behavior near large voids.


Overall, both the strain energy and the chemical binding energy/surface energy contribute to the solute distribution, as clearly illustrated in the Mg-Y system (Fig.~\ref{fig:segregation}).  When the cluster is small, the strain energy effect dominates, causing Y solute atoms to segregate near vacancy clusters.  When the vacancy cluster size increases, the tension stress first increases then decreases (Fig.~\ref{fig:causes}(a)).  Eventually the solute-vacancy binding starts to dominate and cause the depletion of the Y solute.  For Mg-Al, the binding between vacancy and Al is so strong that the depletion due to strain energy was not observed even for a 13-mer that produces the largest hydrostatic stresses under the simulation conditions.  Finally, the Mg-Zn system shows that the 2NN binding cannot be ignored in solute-vacancy interaction.


\begin{figure}
\includegraphics[width=0.48\textwidth]
{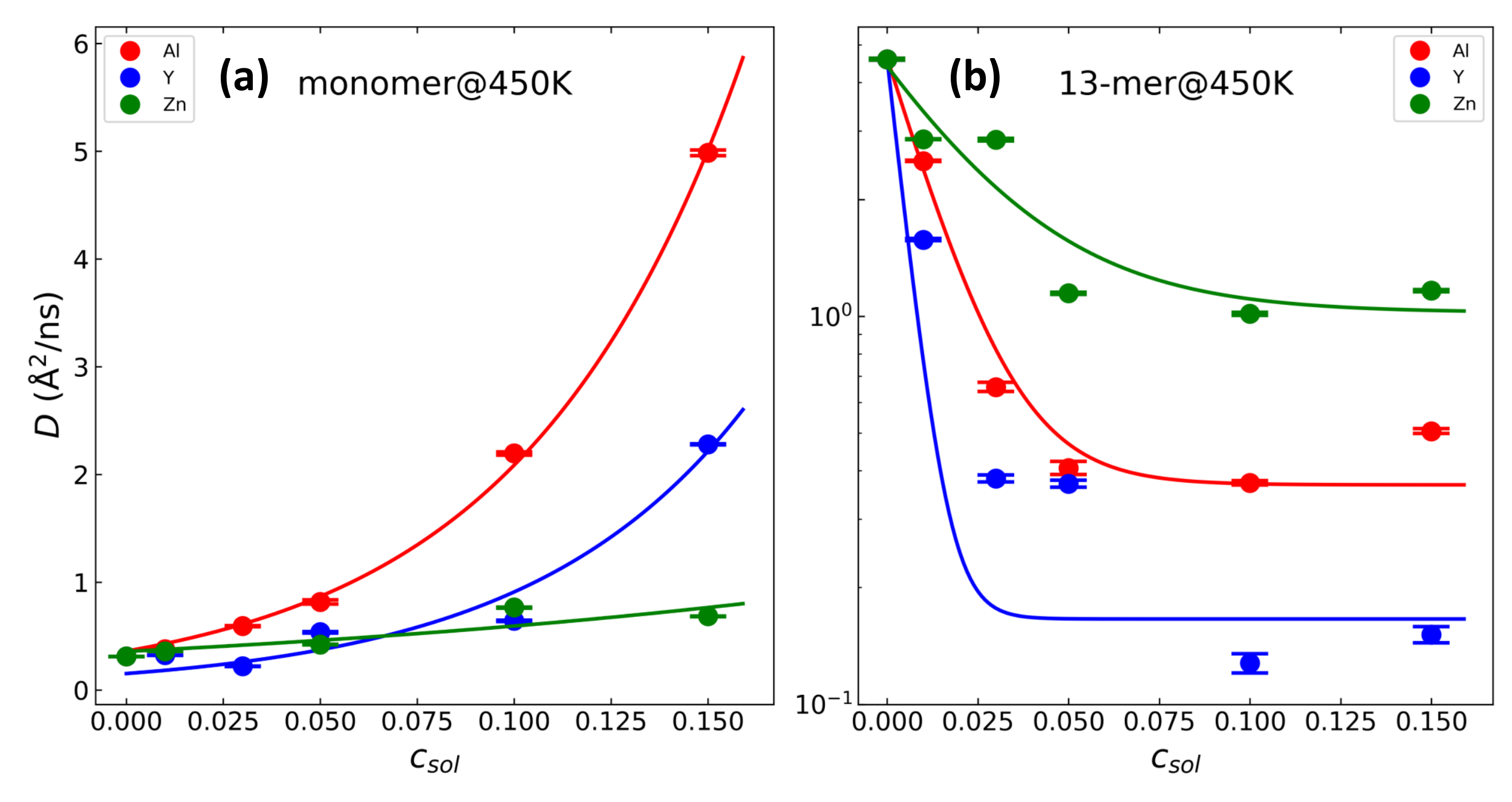}
\caption{\label{fig:diffusivity} Solute concentration dependence of the diffusion coefficients of (a) single vacancy and (b) 13-mer in random solid solutions at 450K. Lines are fitting results.  The fitting formulae are (a) $A\exp(Bc)$ and (b) $A-B(1-\exp(Cc))$, where $A$, $B$, $C$ are fitting parameters and $c$ is solute concentration.}
\end{figure}

On one hand, vacancy clusters attract solute atoms; on the other hand, the solute atoms can stabilize vacancy clusters.  Since solutes are much less mobile than vacancies, the solute-vacancy interaction results in a drag effect on vacancies.  In other words, the solute atoms can trap vacancies and vacancy clusters to extend their lifetimes.  For example, vacancies and vacancy clusters were found to be stabilized by Mg solute atoms in Al alloys in experiments~\cite{Somoza2002}, and by H, Sn, and Nb atoms in Zr alloys in computational studies~\cite{Varvenne2016,Wu2021}.  

\begin{figure*}
\includegraphics[width=0.99\textwidth]{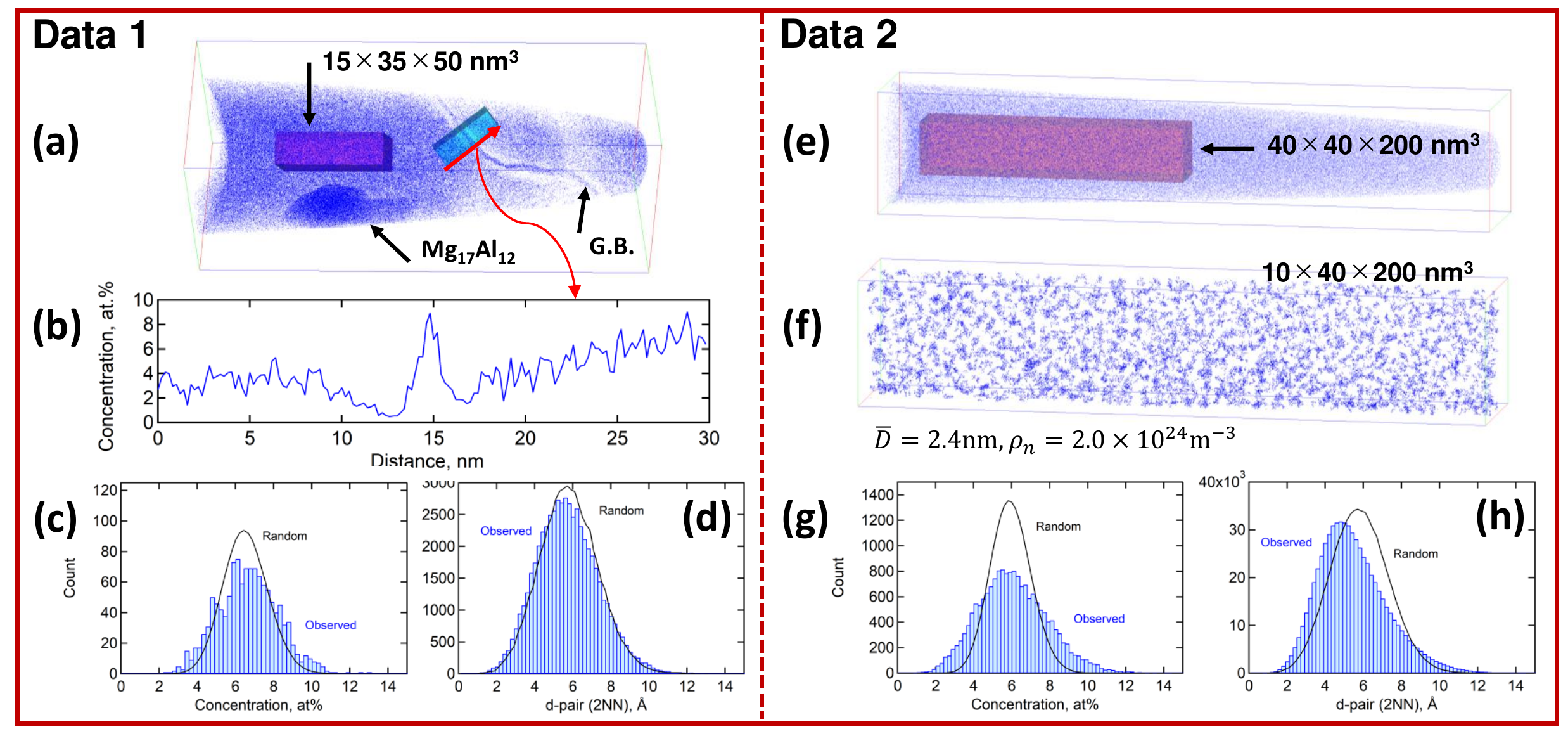}
\caption{\label{fig:exp} (a)-(d) Data 1, (e)-(h) Data 2. (a) APT map of a needle sample (Data 1) from an as-extruded Mg-9wt.\%Al alloy, showing a grain boundary and a Mg$_{17}$Al$_{12}$ precipitates. (b) 1D composition profiles across the grain boundary, position, and direction as shown in the red arrow in (a). (c) Histogram of the composition in blocks of 450 atoms in Data 1. (d) Histogram of the solute 2NN pair distance in Data 1. 
(e) APT map of a second needle sample (Data 2) from the same Mg-Al alloy where Data 1 were extracted. (f) solute clusters identified in a slice of Data 2 using the maximum separation method.  (g) Histogram of the composition in blocks of 500 atoms in Data 2. (d) Histogram of the solute 2NN pair distance in Data 2.}
\end{figure*}

To demonstrate this stabilization effect, we calculated the solute effects on the diffusion coefficients of single vacancies and vacancy clusters in Mg alloys, as shown in Fig.~\ref{fig:diffusivity}.  All three types of solute facilitate the diffusion of single vacancies, regardless of their tendency to bind to a vacancy.  On the other hand, all three solutes hinder the diffusion of 13-mers.  The two different trends are likely related to the transition of the migration mechanism for vacancy clusters, from the exchange mechanism for single vacancies to the surface mechanism for large vacancy clusters.  The combined effect of the increased single vacancy diffusivity and the reduced vacancy cluster diffusivity is that the vacancy clusters may form more quickly and be less mobile.  The trapped vacancy clusters may absorb more vacancies and, in turn, attract more solute atoms.  This positive feedback loop leads to greater stability of vacancy clusters.


To generate excess vacancies that can facilitate solute segregation co-currently, we performed simultaneous deformation and aging to Mg-9 wt.\%Al alloys using ECAE. APT maps of the Al solute atoms in the ECAE'ed sample are shown in Fig.~\ref{fig:exp}.  The two needle samples (Data 1\&2) were taken from the same ECAE'ed sample, a few micrometers apart. The first sample includes a grain boundary and a precipitate (Fig.~\ref{fig:exp}(a)).  Fig.~\ref{fig:exp}(b) shows a solute free zone near the grain boundary, similar to the well known precipitate free zone.  The measured bulk solute concentration shown in Fig.~\ref{fig:exp}(b) is lower than that with which the samples were solutionized because a large portion of solute atoms have been collected into precipitates.  Fig.~\ref{fig:exp}(c)(d) show the histogram of the composition in blocks of 450 atoms, as well as that of the 2NN pair distance between solute atoms.  They resemble those of the random solid solution, indicating negligible solute clustering.  

The second sample is away from the grain boundary or other visible defects (Fig.~\ref{fig:exp}(e)).  The solute clusters identified using the maximum separation method~\cite{VAUMOUSSE2003,HYDE2011} are shown in Fig.~\ref{fig:exp}(f).  The clusters average 2.4nm in diameter with a number density of $2\times10^{24} \mathrm{m}^{-3}$.  The average solute concentration within the clusters is 14.8 at.\%, which is significantly higher than the composition in the matrix, about 6 at.\% (Fig.~\ref{fig:exp}(g)).  The clustering is also shown by the histograms of the composition and the 2NN pair distance, both of which  significantly deviate from those of the random solid solution (Fig.~\ref{fig:exp}(g)(h)).  

Homogeneous solute clustering in Mg-Al alloys (Mg-rich) is unstable according to the free energy landscape, as opposed to the Al-Mg alloys (Al-rich)~\cite{Epler2004,Kleiven2019}.  As a result, there were no reported observations of GP zones or solute segregation in Mg-Al alloys statically aged~\cite{Nie2012}.  Therefore, the solute segregation clustering observed in Data 2 must be driven by local heterogeneities generated by deformation.  Data 1 and 2 are from the same ECAE'ed sample, except that there is a grain boundary in Data 1.  Since grain boundaries are vacancy sinks, the lack of solute segregation in Data 1 is most likely associated with the depletion of vacancies near the grain boundary.  


In conclusion, we used atomistic simulations to demonstrate vacancy cluster induced solute segregation/depletion and that the solute atoms can in turn stabilize vacancy clusters.  This interplay can lead to an extended lifetime of vacancy clusters to serve as heterogeneous sites for solute segregation and intermetallic precipitation. Our experimental observation in dynamics aged Mg-Al alloys of solute segregation, which is absent in statically aged Mg-Al alloys, provides strong support to this mechanism.  Further experimental characterization of vacancy clusters using, for example, a combination of positron annihilation lifetime spectroscopy (PALS), Atomic Electron Tomography (AET), and high-resolution TEM, is crucial in developing a quantitative model for this mechanism in guiding the design of deformation assisted processing.

\begin{acknowledgments}

The authors thank the helpful discussion with Professor Kazuhiro Hono.  P. Yi, S.E. Prameela, T.P. Weihs, and M.L. Falk acknowledge the support from the Army Research Laboratory accomplished under Cooperative Agreement Number W911NF-12-2-0022.  The views and conclusions contained in this document are those of the authors and should not be interpreted as representing the official policies, either expressed or implied, of the Army Research Laboratory or the U.S. Government.  The U.S. Government is authorized to reproduce and distribute reprints for Government purposes notwithstanding any copyright notation herein.  Computational resources from Maryland Advanced Research Computing Center (MARCC) are acknowledged.  T. Sasaki acknowledges the support by JSPS KAKENHI (Grant Number JP21H01675) and Element Strategy Initiative of MEXT (Grant Number JPMXP0112101000). 
\end{acknowledgments}

\appendix

\section{Supplementary Materials}

\subsection{Simulation methods}

Molecular dynamics (MD) and Monte Carlo (MC) simulations were performed using the LAMMPS package~\cite{Plimpton1995}.  We chose the MEAM potentials to model three Mg-X binary alloy systems, where X=Al, Y, and Zn~\cite{Kim2009,Kim2015,Jang2018}.  We have used these systems in our previous studies on dislocation and twin mobilities~\cite{Yi2016, Yi2017a, Yi2019, Yi2021}.  

Hybrid MC/MD simulations were used to sample the equilibrium solute distribution near vacancy clusters.  The Monte Carlo moves involve atom-type swaps for randomly chosen solvent-solute atom pairs, and the attempts were accepted or rejected based on the energy change due to the swap~\cite{Sadigh2012, Yi2017b, Yi2018, Yi2020}.  The system contains about 40000 atoms.  It is orthorhombic and each dimension is about 100$\mathrm{\AA}$ long.  Periodic boundary conditions were applied in all three directions. 

We additionally deployed density functional theory (DFT) calculations to validate the accuracy of the predictions made with these MEAM potentials.  The DFT calculations were performed using the VASP package~\cite{Kresse1996a, Kresse1996b} with the projector augmented wave method (PAW)~\cite{Blochl1994b}.  The electron exchange and correlation energies were calculated using the Perdew-Berke-Ernzerhof (PBE) generalized gradient approximation (GGA)~\cite{Perdew1996}.  The energy cutoff was 350eV.  To calculate energy, we first performed volume optimization with energy and force tolerances of $10^{-5}$eV and 0.01eV/$\mathrm{\AA}$, respectively.  Once the volume was optimized and fixed, the energy was converged to within $10^{-6}$eV for more accurate energy calculations.  The tetrahedron method with Bl{\"{o}}chl corrections~\cite{Blochl1994a} was applied for these energy calculations.  The Monkhorst-Pack $\Gamma$-centered $k$-point mesh was used~\cite{Monkhorst1976}.  The supercell consists of $6\times6\times3$ hexagonal unit cells ($6\times6$ on the basal plane, and 3 along the $c$-axis) and 216 lattice sites, and the $k$-point mesh is $8\times8\times8$.

The solute-vacancy binding energy was calculated using the formula~\cite{Shin2010},

\begin{eqnarray}
 -E_{binding}=& &E(\mathrm{Mg}_{\mathrm{N}-2}\mathrm{X}_1\square_1)+E(\mathrm{Mg}_N)\nonumber\\
  & &-E(\mathrm{Mg}_{N-1}\mathrm{X}_1)-E(\mathrm{Mg}_{N-1}\square_1),
\label{eq:one}
\end{eqnarray}

\noindent where $\mathrm{X}$ and $\square$ represent a solute atom and a vacancy in the Mg matrix, respectively.  The minus sign is to keep the convention that a positive binding energy indicates attraction.  The energies were calculated at 0K.  The solute positions are the 12 nearest neighbor (NN) sites (distance to origin $<$ 3.5$\mathrm{\AA}$) and the 44 second nearest neighbor (2NN) sites (3.5$\mathrm{\AA}$ $\leq$ distance to origin $<$ 7$\mathrm{\AA}$).  Considering the symmetry, only 2 of the 12 NN sites and 6 of the 44 2NN sites are unique. 

The MD simulations used a simulation system of 11520 atoms in a simulation cell approximately 60$\rm{\AA}$ on a side.  For DFT calculations, a supercell of 216 lattice sites was used.  This is larger than some of the earlier DFT studies~\cite{Shin2010, Zhou2016}.  During the calculation, system volume, shape, and atom positions were all allowed to relax.  However, the hexagonal symmetry is retained.  In some studies, lattice parameters determined in pure Mg were used throughout~\cite{Huber2012}.

When the vacancy clusters grow in size, the surface energy becomes a more relevant quantity than the solute-vacancy binding energy.  The surface energy was calculated using the usual slab method,  

\begin{equation}
   \gamma=(E_{slab}-NE_{bulk})/2A,
\label{eq:gammaf}
\end{equation}

\noindent where $E_{slab}$ is the total energy of the relaxed surface slab, $N$ is the number of unit layers in the slab, $E_{bulk}$ is the total energy of the bulk unit layer, and $A$ is the surface area.  

We first relaxed the bulk system to optimize lattice parameters, maintaining the hexagonal symmetry.  During this step we also calculated the values of $A$ and $NE_{bulk}$.  Then we fixed the lattice parameters and created a slab by adding a vacuum layer 12$\mathrm{\AA}$ thick.  Next, we relaxed the atom positions and computed $E_{slab}$.  During this step, the supercell dimensions were fixed.  Finally, the value of $\gamma$ was obtained using Eq.~(\ref{eq:gammaf}).

The same procedure was applied to solid solutions, with one or more atoms in the matrix randomly chosen and replaced with substitutional solute atoms.  We calculated the surface energies for two planes, the basal plane (11-20) and the prismatic plane (1-100).  Ten randomized atomic configurations of each solute concentration were used to collect statistics.  Typically, only one unit cell on the surface is needed for calculating the surface energy of the pure metal.  When calculating the surface energy for a solid solution, a unit layer contains multiple unit cells on the surface.  The crystal has a dimension of about 20$\times$20$\times$50 $\mathrm{\AA}^3$, which is chosen to ensure that the finite size effect is negligible.

Lastly, to track the locations of the diffusing vacancies or vacancy clusters, we compared the structure of each snapshot to a reference perfect crystal during post-analysis and identified the missing atoms as vacancies.  This comparison was performed after removing the thermal fluctuations by performing a short conjugate gradient energy minimization with an energy tolerance of 10$^{-12}$. 

\subsection{Experiments}

An as-cast binary Mg-9wt.\%Al alloy was purchased from Magnesium Elektron North America (MENA), Madison, IL.  The as-received ingots were found to be randomly textured with precipitates within the bulk~\cite{Ma2019}.  To create and start with a precipitate-free and well-controlled material, the as-received ingots were subjected to solution-treatment at 450\textdegree C for 24 h within a protective argon gas flow, followed by quenching in ice water to avoid precipitation.  Rectangular-shaped billets with a size of 6.35$\times$6.35$\times$10 mm$^3$ were cut from the center of the solution-treated sheet for ECAE processing.  The cut samples were subjected to the one-pass ECAE at 150\textdegree C with a right-angle die.  The extrusion rate was 0.15 mm/min, and a backpressure of 0.45 MPa was applied.  One pass takes approximately 2 h to complete.  To characterize the samples, APT mapping was performed using a local electrode atom probe (LEAP 5000XS), in voltage pulse mode at a temperature of 30 K.  The needle-shaped samples for the 3DAPT analysis were prepared by the standard lift-out technique using FEI Helios G4 UX. 

Solute clusters and precipitate particles were identified from APT images using the maximum separation method~\cite{VAUMOUSSE2003,HYDE2011}.  The parameters were chosen to be $D_{max}$=5 $\mathrm{\AA}$, $N_{min}$=10, $L$=6.8 $\mathrm{\AA}$, and $D_{erosion}$=3.4 $\mathrm{\AA}$.



%

\end{document}